\documentclass{ws-ijmpb}
\usepackage{graphicx}
\usepackage{epsf,latexsym}
\usepackage{times}

\newcommand{\ket}[1]{\ensuremath{\left| #1 \right\rangle}}

 \newcommand{\EX}[1] {\ensuremath{\left\langle #1 \right\rangle}}
 
\newcommand{\ihbar}{\ensuremath{\frac{i}{\hbar}}}
\newcommand{\half} {\ensuremath{\frac{1}{2}}}

\newcommand{\be}{\begin{equation}}
\newcommand{\ee}{\end{equation}}
\newcommand{\ba}{\begin{eqnarray}}
\newcommand{\ea}{\end{eqnarray}}

\newcommand{\BUE}{Centre for Theoretical Physics,
                  The British University in Egypt,
                  El Sherouk City, Postal No. 11837, P.O. Box 43, Egypt.}

\newcommand{\LBRO}{Department of Physics, Loughborough University,
                   Loughborough, Leics LE11 3TU, United Kingdom}
\begin{document}


\title{Non-linear dynamics, entanglement and the quantum-classical crossover of two coupled SQUID rings}

\author{M. J. Everitt}
\address{\BUE\\ \LBRO\\ m.j.everitt@physics.org}

\maketitle
\begin{history}
\received{Day Month Year}
\revised{Day Month Year}
\end{history}

\begin{abstract}
  We   explore  the  quantum-classical   crossover  of   two  coupled,
  identical,  superconducting   quantum  interference  device  (SQUID)
  rings.  We  note that  the motivation  for this work  is based  on a
  study   of  a   similar  system   comprising  two   coupled  Duffing
  oscillators.   In   that  work  we  showed   that  the  entanglement
  characteristics  of  chaotic   and  periodic  (entrained)  solutions
  differed significantly and that  in the classical limit entanglement
  was preserved only in  the chaotic-like solutions.  However, Duffing
  oscillators are a highly idealised toy model. Motivated by a wish to
  explore  more experimentally  realisable systems  we now  extend our
  work to  an analysis  of two coupled  SQUID rings.  We  observe some
  differences in behaviour  between the system that is  based on SQUID
  rings rather than on Duffing oscillators.  However, we show that the
  two systems  share a  common feature. That  is, even when  the SQUID
  ring's  trajectories  appear   to  follow  (semi)  classical  orbits
  entanglement persists.  
\end{abstract}


\maketitle

\section{Introduction}

In  this paper  we report  the results  of an  initial study  into the
entanglement   properties  associated   with  the   quantum  classical
crossover of two coupled SQUID rings. This work can be
considered as being  part of a much larger study spanning many
years                             (see,                            for
example,~\refcite{bha1,bha2,gho1,sch1,Per98,Hab98,Spiller95,Greenbaum07,brun97,Bru96,Spi94,Gis93,Gis93b,everitt05,everitt05b,Everitt07,everitt0708}). The
overarching  theme being  the quantum-classical
crossover  of  systems  the   exhibit  non-linear  dynamics  in  their
classical   motion.  Our   interest   arises  from   the  fact   the
Schr\"odinger   equation   is  linear.   Hence,   when  applying   the
correspondence   principle,  how  can   one  recover   classical  like
trajectories of classically non-linear systems in terms of expectation
values of  quantum mechanical observables? A  general solution appears
to  take the  form  of introducing  environmental  degrees of  freedom
together with a suitable expression of the correspondence principle.

This    particular    study    was    motivated    by    a    previous
investigation\cite{everitt05} of two  coupled duffing oscillators. In
that work  we wished to extend  the analysis of  the quantum classical
crossover           of            a           single           Duffing
oscillator\cite{bha1,bha2,gho1,sch1,Per98,Hab98,brun97,Bru96} to more
complex systems.  Specifically we wished  to study the the recovery of
classical  like  orbits from  a  quantum  description  of two  coupled
identical Duffing oscillators.  Here our interest lay in understanding
how  the purely  quantum  mechanical phenomena  of entanglement  would
change  as the  coupled  system approached  the  classical limit.   We
showed   \emph{``that   the   entanglement  characteristics   of   two
  ‘classical’   states  (chaotic   and   periodic  solutions)   differ
  significantly  in the  classical limit.  In particular,  we show[ed]
  that significant  levels of entanglement  are preserved only  in the
  chaotic-like  solutions''}\cite{everitt05}.   Although the  Duffing
oscillator has been studied for  many years with great interest it is,
nevertheless, a  toy model.  Hence, recently, our  focus has  moved to
include the study of SQUID  rings. These are devices that can manifest
similar properties  as the Duffing  oscillator (such as a  double well
potential  and classical,  dissipative,  chaos) but  are also  readily
fabricated.  Hence,  in  this  paper  we  investigate  the  entanglement
characteristics  for   the  quantum-classical  crossover   of  a  system
comprising of two identical coupled SQUID rings.

The correspondence principle in quantum mechanics is usually expressed
in  the  form: \emph{``For  those  quantum  systems  with a  classical
  analogue,  as  Planck's   constant  becomes  vanishingly  small  the
  expectation  values  of  observables  behave  like  their  classical
  counterparts''}\cite{Mer98}.   Applying  this  expression   of  the
correspondence principle  together with introduction  of environmental
degrees of freedom via unravelling the master equation is the standard
method   of   recovering   trajectories  of   classically   non-linear
systems.  Indeed, we  used this  procedure to  recover  classical like
trajectories for our work on two coupled Duffing oscillators. However,
in~\refcite{everitt0708}   we  found   that  this   formulation   of  the
correspondence principle cannot be  applied to SQUID rings. Instead we
proposed a more pragmatic version as follows:
Consider $\hbar$ fixed (it is) and scale the Hamiltonian so that when
compared with the minimum area $\hbar/2$ in phase space:
\begin{itemize}
  \item[(a)] the relative motion of the expectation values of the 
	           observable become large and
  \item[(b)] the state vector is localised.
\end{itemize}
Then, under these circumstances, expectation values of operators will
behave like their classical counterparts\cite{everitt0708}.

In  this  work   the  the  evolution  of  the   state  vector  $\left|
\psi\right\rangle$ of the open  system comprising of two coupled SQUID
rings subject to Ohmic damping  will be modelled using a quantum state
diffusion  unravelling of the  master equation.  This model  takes the
form of an It\^{o} increment equation given by\cite{Gis93,Gis93b}
\begin{eqnarray}\label{eq:qsd}
  \ket{d\psi}  &  =&-\ihbar \hat{H}_{sys} \ket{\psi} dt\nonumber\\
   &&  +\sum_{j}\left[  \EX{\hat{L}_{j}^{\dagger}} \hat{L}_{j}-\half \hat{L}_{j}^{\dagger}\hat{L}_{j}-\half   
   \EX{\hat{L}_{j}^{\dagger}} \Bigl\langle \hat{L}_{j}\Bigr\rangle \right]  \ket{\psi} dt\nonumber\\
   &&  +\sum_{j}\left[  \hat{L}_{j}-\EX{\hat{L}_{j}} \right]  \ket{\psi} d\xi
\end{eqnarray}
where the  first term on the right  hand side of this  equation is the
Schr\"odinger    evolution   of    the    system   with    Hamiltonian
$\hat{H}_{sys}$.  The effect  of the  environment is  modelled  by the
addition  of  a  second  (drift)  and third  (fluctuation)  term  that
describe  the decohering  effects of  the environment  on  the systems
state vector. In this work  we consider only one Lindblad operator per
subspace    $\hat{L_j}=\sqrt{2\zeta}\hat{a_j}$,    where   $a_j$    is
annihilation operator.  The time increment  is $dt$ and the $d\xi$ are
complex             Weiner             increments            satisfying
$\overline{d\xi^2}=\overline{d\xi}=0$        and       $\overline{d\xi
  d\xi^{*}}=dt$\cite{Gis93,Gis93b}  where  the  over-bar denotes  the
average over infinitely many stochastic processes.

\section{Review of the entanglement properties of two coupled Duffing oscillator}
In order to set the scene for  our new results we now summarise some of
the text and results from  our earlier work on entanglement dynamics in
coupled Duffing oscillators\cite{everitt05}. In that work we extended
the analysis  of previous work on  such systems\cite{Bru96,Per98} and
considered  two  identical,  coupled  Duffing oscillators.  The
Hamiltonian for each oscillator was given by
\begin{equation}
H_{i}=\frac{1}{2}p_{i}^{2}+\frac{\beta^{2}}{4}q_{i}^{4}-\frac{1}{2}q_{i}
^{2}+\frac{g_{i}}{\beta}\cos\left(  t \right)  q_{i}+\frac
{\Gamma_{i}}{2}(q_{i}p_{i}+p_{i}q_{i})
\end{equation}
where $q_{i}$ and $p_{i}$ are  the position and momentum operators for
each oscillator.   As usual for  Ohmic damping, the  Lindblad operators
were  simply  $L_{i}=\sqrt{2\Gamma_{i}}  a_{i}$ (for  $i=1,2$),  where
$a_{i}$ is  the annihilation operator.   As with previous work  in this
field\cite{Bru96,Per98}  in~\refcite{everitt05}  we  set the  parameters
$g_{i}=0.3$ and $\Gamma_{i}=0.125$  respectively.  The Hamiltonian for
the coupled system then was taken to be:
\begin{equation}\label{eq:hamSys}
H=H_{1}+H_{2}+\mu q_{1}q_{2}
\end{equation}
where  we  set the  coupling  strength  $\mu=0.2$.  In this  work  the
parameter $\beta$  is a scaling  parameter for $\hbar$  with $\beta=1$
being the fully classical limit.

\begin{figure}[!tb]
\centerline{\resizebox*{1.0\textwidth}{!}{\includegraphics{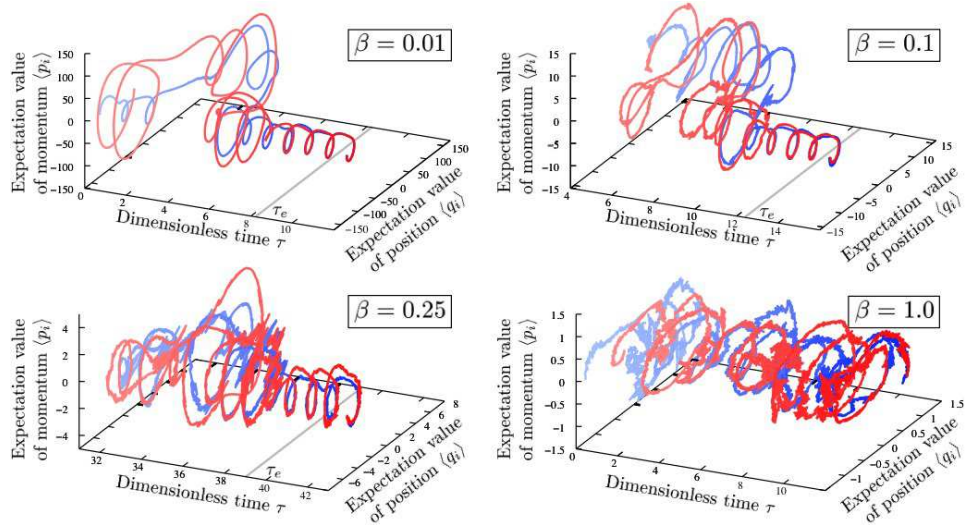}}}
\vspace*{8pt}
\caption{\label{fig:dyn}The dynamics of the expectation values of
  position and momentum as a function of time (normalised to drive
  periods) for $\beta=0.01$, $\beta=0.1$, $\beta=0.25$ and
  $\beta=1.0$.  The dynamics have been taken over the same duration
  and displayed so that, with the exception $\beta=1.0$, they
  approximately align at the time $(\tau_e)$ at which entrainment
  occurs. (NB: Figure and caption reproduced
  from~14)}
\end{figure}
Solutions   to  the   equation  of   motion  for   the   state  vector
(\ref{eq:qsd})  are shown  in Figure~\ref{fig:dyn}.  Here we  show the
evolution of the  expectation values of position and  momentum for the
two  coupled oscillators as  a function  of time  for a  four different
values of $\beta$.  The  emergence of classical trajectories for small
$\beta$  is clearly  apparent. Furthermore  for $\beta=0.01,  0.1$ and
$0.25$ chaotic behaviour of the  ceases after some time and the motion
becomes almost periodic.

We measured  the entanglement between  the two oscillators  by computing
the entropy  of entanglement for the system\cite{Ben96}.  That is the
von  Neumann  entropy\cite{Nie00}  of  the reduced  density  operator
$\rho_i$ for one of the oscillators i.e.
$$
S\left(\rho_{i}\right)
=-\mathrm{Tr}[\rho_{i}\ln\rho_{i}].
$$ 

\begin{figure}[!tb]
\centerline{\resizebox*{0.8\textwidth}{!}{\includegraphics{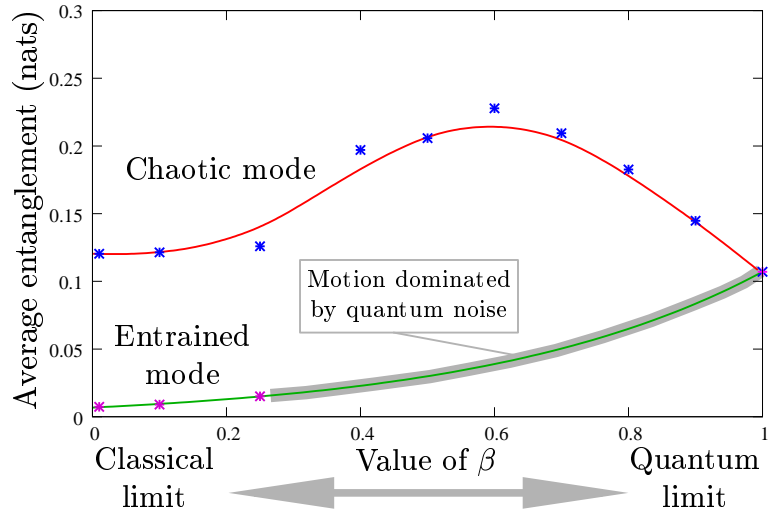}}}
\vspace*{8pt}
\caption{\label{fig:ent}Mean  entanglement entropy  as  a function  of
  $\beta$ for  the chaotic-like and periodic  (entrained) states. Here
  we see that the entanglement entropy for system in the chaotic state
  does  not vanish as  $\beta$ approaches  the classical  regime. (NB:
  Figure              and              caption              reproduced
  from~14)\label{fig:average.qsd}}
\end{figure}
We then demonstrated that  that when these coupled Duffing oscillators
enter  the  almost periodic  entrained  state  the entanglement  falls
rapidly as the system  approaches the classical regime.  However, when
both oscillators undergo chaotic like motion we found that significant
average entangled remained even when the dynamics of the system appear
to be  classical. We found these results (shown in figure~\ref{fig:ent}) surprising, as entanglement,
which  does  not have  a  classical  counterpart,  persisted even  for
classical looking trajectories.

\section{RSJ model and scaling the dynamics}
As already  noted, in~\refcite{everitt0708} we found that  the SQUID ring
Hamiltonian was incompatible  with the correspondence principle phrased
in terms  of making Planck's  constant vanishingly small. A  solution to
this problem  was found  by scaling the  system so that  the underlying
dynamics remain,  qualitatively, the same. We  can see how  this may be
achieved  for SQUID  rings by  looking  at the  classical equations  of
motion that are given by the resistively shunted junction (RSJ) model.
The RSJ  equation of  motion for the  magnetic flux, $\Phi$,  within a
driven SQUID ring is\cite{Likharev86}:
\begin{equation}\label{eq:rsj}
C \frac{d^2\Phi}{dt^2}+\frac{1}{R}\frac{d\Phi}{dt}+\frac{\Phi-\Phi_x}{L}+I_c \sin\left(\frac{2\pi \Phi}{\Phi_0}\right)
=I_d \sin \left(\omega_d t \right)
\end{equation}
where  $\Phi_x$,  $C$,  $I_c$,  $L$  and $R$  are,  respectively,  the
external flux bias, capacitance and critical current of the weak link,
the inductance of the ring and the resistance. The drive amplitude and
frequency are  $I_d$ and $\omega_d$ respectively  and $\Phi_0=h/2e$ is
the flux quantum.

We  can  then  rewrite   (\ref{eq:rsj})  in  the  standard,  universal
oscillator   like,   form  by   making   the  following   definitions:
$\omega_0=1/\sqrt{LC}$,               $\tau=\omega_0               t$,
$\varphi=(\Phi-\Phi_x)/\Phi_0$, $\varphi_x=\Phi_x/\Phi_0$, $\beta=2\pi
L  I_c/\Phi_0$,  $\omega=\omega_d/\omega_0$, $\varphi_d=I_d  L/\Phi_0$
and   $\zeta=1/2\omega_0    RC$. This yields the following equation of motion:
\begin{equation}\label{eq:rsjNorm}
\frac{d^2\varphi}{d\tau^2}+2\zeta\frac{d\varphi}{d\tau}+\varphi+\frac{\beta}{2\pi} \sin\left[2\pi\left( \varphi+\varphi_x\right)\right]
=\varphi_d \sin \left(\omega \tau \right)
\end{equation}
In  this system of  units we  then see  that we  can scale  the system
Hamiltonian  through   changing  either  $C  \rightarrow   aC$  or  $L
\rightarrow  bL$  so long  as  we  also  make the  following  changes:
$R\rightarrow\sqrt{{b}/{a}}R$,  $I_d\rightarrow  {I_d}/{\sqrt{b}}$ and
$\omega_d\rightarrow {\omega_d}/{\sqrt{ab}}$.

In  this   paper  we  use  the  following   basic  circuit  parameters
$C=1\times10^{-13}$F,  $L=3\times10^{-10}$H, $R=100\Omega$, $\beta=2$,
$\omega_d=\omega_0$ and $I_d=0.9\,\mu\mathrm{A}$. We note that we have
biased the ring at the  half flux quantum, $\Phi_x=0.5\Phi_0$, so that
the potential  approximates a double well.  We change $a$  so that $C$
varies  between  $1\times 10^{-16}$\,F  (quantum  limit) and  $1\times
10^{-9}$\,F  (classical  limit),  changing  other  circuit  parameters
in line with the above methodology.

\section{Quantum mechanical description of the SQUID Ring}
We now turn to the quantum  description of this system. The SQUID ring
Hamiltonian   for  each  of   the  identical   rings  on   their  own
is\cite{Barone1982}:
\begin{equation}\label{eq:qmBase}
\hat{H_i}=\frac{\hat{Q}^2_i}{2C}+\frac{\left(\hat{\Phi}_i-\Phi_{x_i}(t)\right)^2}{2L}-
\frac{\hbar I_c}{2e} \cos\left(\frac{2\pi\hat{\Phi}_i}{\Phi_0} \right)
\end{equation}
where the  magnetic flux threading each ring,  $\hat{\Phi}_i$, and the
total charge  across each weak link  $\hat{Q}_i$ take on  the roles of
conjugate  variables  for  the  system with  the  imposed  commutation
relation        $\left[\hat{\Phi}_i,\hat{Q}_i\right]=i\hbar$       and
$\Phi_0=h/2e$  is  the  flux  quantum.  Here  $\Phi_{x_i}(t)$  is  the
external  applied magnetic flux  and incorporates  the drive  term for
each ring.

We define dimensionless flux and  charge operators in the manner usual
for          the          simple         harmonic          oscillator:
$\hat{x}_i=\sqrt{{C\omega_0}/{\hbar}}\hat{\Phi}_i$                  and
$\hat{p}_i=\sqrt{{1}/{\hbar C \omega_0}}\hat{Q}_i$.

$\hat{H}_i'=\hat{H}_i/\hbar \omega_0$ we find that
\begin{equation}\label{eq:qmNorm}
  \hat{H}_i'=\frac{\hat{p}^2_i}{2}+\frac{[\hat{x}_i-x_i(t)]^2}{2}-\frac{I_c}{2e \omega_0}\cos \left( \Omega \hat{x}_i\right)
\end{equation}
where $\Omega=\left[(4e^2/\hbar)\sqrt{(L/C)} \right]^{1/2}$.

We note that  this method of introducing Ohmic  damping does not bring
with it  the frequency shift that  arises through the  damping term in
the classical  dynamics. We resolve  this problem, as for  the Duffing
oscillator,   by   the   addition    of   an   extra   term   to   the
Hamiltonian\cite{Mer98,Per98}. That is, the Hamiltonian
\begin{equation}\label{eq:qmNorm2}
  \hat{H}'_i=\frac{\hat{p}^2_i}{2}+\frac{[\hat{x}_i-x_i(t)]^2}{2}-\frac{I_c}{2 e \omega_0}\cos \left( \Omega   
           \hat{x}_i\right)+\frac{\zeta}{2}\left(\hat{p}_i\hat{x}_i+\hat{x}_i\hat{p}_i\right)
\end{equation}
is one we can use to generate trajectories that will be comparable to those predicted by the RSJ model.
\begin{eqnarray}\label{eq:HamCoupled}
  \hat{H}_{total}&=&\sum_{i\in\{1,2\}}\left\{\frac{\hat{p}^2_i}{2}+\frac{[\hat{x}_i-x_i(t)]^2}{2}-\frac{I_c}{2 e \omega_0}\cos \left( \Omega   
           \hat{x}_i\right)+\right. \nonumber \\ 
           &&\left.\frac{\zeta}{2}\left(\hat{p}_i\hat{x}_i+\hat{x}_i\hat{p}_i\right)\right\}+\mu \hat{x}_1 \hat{x}_2 \nonumber
\end{eqnarray}
where $\mu=0.2$

\section{Results and comparison with the Duffing oscillator}
\begin{figure}[!tb]
\centerline{\resizebox*{0.8\textwidth}{!}{\includegraphics{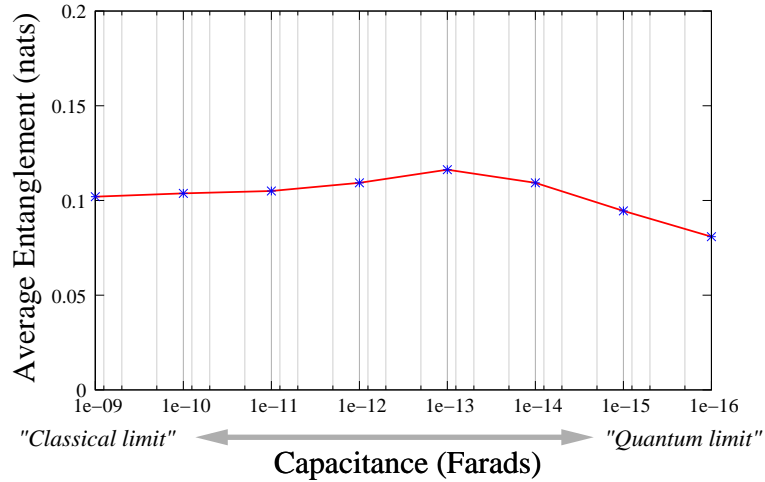}}}
\vspace*{8pt}
\caption{Mean entanglement  entropy as  a function of  Capacitance two
  coupled SQUID rings.  Here we  see that the entanglement entropy for
  system  does  not  vanish   even  as  it  approaches  its  classical
  limit.\label{fig:s1}}
\end{figure}
We are  now in  a position  to present results  for two  coupled SQUID
rings for  comparison with figure~\ref{fig:average.qsd}.  Here we have
computed the average entanglement as  a function of capacitance (for a
selection  of  scale  parameters   $a$).  However  we  note  that  the
entanglement entropies  presented here  are is the  average entanglement
over either a long time period or many similar trajectories. It is not
the entanglement associated with  the average density operator taken of
many  experiments.   This  average  entanglement   cannot  therefore  be
considered   usable    in   a   quantum    information   sense.    In
figure~\ref{fig:s1} we show this average entanglement entropy. Here the
averaging  of each  trajectory  was  determined on  a  point by  point
basis.  A sufficient  averaging  was used  so  as to  ensure that  the
results presented here had settled to within a percent or so.

As for  the Duffing oscillators,  here the mean entanglement  does not
appear to  vanish in the classical limit  (large capacitance). Another
surprising  feature in common  with the  Duffing oscillator  results is
that  the average entropy  is not  maximum at  the most  quantum limit
(smallest capacitance).

\begin{figure}[!tb]
\centerline{\resizebox*{1.0\textwidth}{!}{\includegraphics{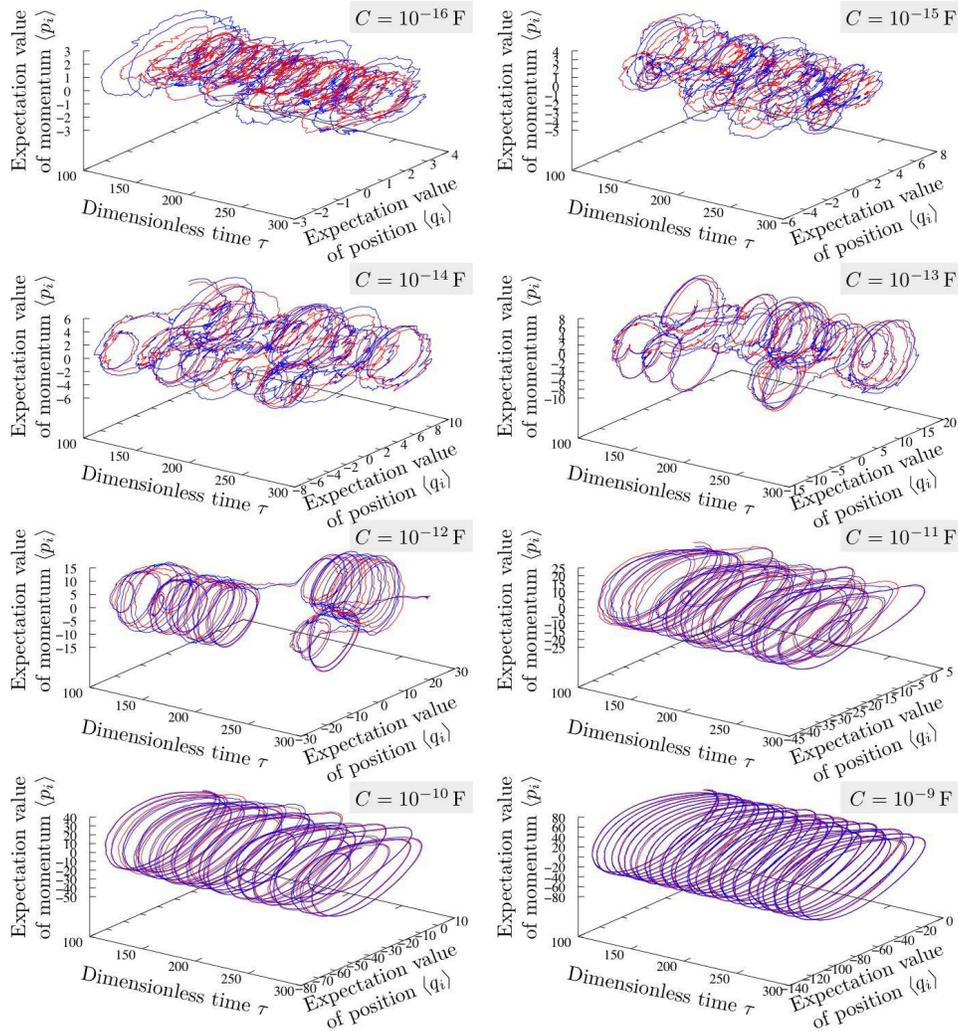}}}
\vspace*{8pt}
 \caption{The dynamics of the expectation values of normalised flux and charge
 as a  function of dimensionless time for a range of capacitances.\label{fig:s2}}
\end{figure}
From  our previous study  of the  Duffing oscillator  as well  as that
presented in~\refcite{Gho04}  we expect this average  entanglement of the
pure state to be associated  with and underlying chaotic or non-linear
dynamics.  In figure~\ref{fig:s2}  we  show sample  phase
portraits for the expectation values  of flux and charge as a function
of dimensionless time.  For very small capacitance's we  see the motion
is dominated  by noise fluctuations.  As the capacitance  increases we
see,   as   expected,    the   emergence   of   non-linear   dynamical
behaviour. However rather surprisingly we  see that for large values of
capacitance (i.e. the classical  limit) this non-linear dynamics gives
way to  a nearly periodic  motion. This result  for SQUID rings  is in
stark  contrast  to  those   for  the  system  comprising  of  Duffing
oscillators.  Whilst we find  these results  most interesting  we are,
here, reporting  our preliminary results  in the study of  coupled SQUID
rings in  the correspondence limit.  We will explore these  results in
more detail in later work.

\section{Conclusion}
We  have  shown   that  for  two  coupled  SQUID   rings  the  average
entanglement   persists  as   the  rings   approach   their  classical
limit. Indeed the trend seems very similar to that associated with the
chaotic motion  of two coupled  Duffing oscillators. We note  that for
the uncoupled rings the dynamics  remains non-linear for all values of
the scale  parameter $a$. However, we  have seen this is  not the case
for  the coupled  system.  Unlike the  entanglement  entropies for  two
coupled Duffing oscillators with the  SQUID ring based system we see no
sudden change  in the mean entanglement entropy associated with this change in the underlying dynamics.  Specifically, even in
nearly  periodic   orbits  the  SQUID   rings  exhibit  non-negligible
entanglement entropies even when  the underlying dynamics appears almost
classical and periodic.  As this  entanglement  is associated  with nearly  periodic
motion it  may form the  basis for extracting usable  entanglement from
this system. This is something that we will explore in a more detailed
investigation of this system at a later date.

\section*{Acknowledgments}
The author  would like  to thank The  Physics Grid  (Loughborough) and
Loughborough University HPC service for use of their facilities.

\bibliographystyle{unsrt}
\bibliography{references}
\end{document}